\documentclass[preprint,preprintnumbers,amsmath,amssymb,superscriptaddress]{revtex4}
\usepackage{txfonts}% Physical Review B
\usepackage{graphicx}% Include figure files
\usepackage{dcolumn}% Align table columns on decimal point
\usepackage{bm}% bold math
\usepackage{array}
\usepackage{wrapfig}
\usepackage[colorlinks=true,plainpages=false,linkcolor=blue,urlcolor=blue,citecolor=blue,pdfpagemode=UseNone,pdfstartview=FitBH]{hyperref}

\begin{document}

\title{Intertwined Charge and Spin Density Waves in Trilayer Nickelate La$_4$Ni$_3$O$_{10}$ Revealed by $^{139}$La NQR}

\author{Jie Dou}
\thanks{These authors contributed equally to this work.}
\affiliation{Institute of Physics, Chinese Academy of Sciences,\\
	and Beijing National Laboratory for Condensed Matter Physics, Beijing 100190, China}
\affiliation{School of Physical Sciences, University of Chinese Academy of Sciences, Beijing 100190, China}

\author{Feiyu Li}
\thanks{These authors contributed equally to this work.}
\affiliation{State Key Laboratory of Crystal Materials, Institute of Crystal Materials, Shandong University, Jinan, Shandong 250100, China}

\author{Mingxin Zhang}
\thanks{These authors contributed equally to this work.}
\affiliation{State Key Laboratory of Quantum Functional Materials, School of Physical Science and Technology, ShanghaiTech University, Shanghai 201210, China}

\author{Jun Luo}
\affiliation{Institute of Physics, Chinese Academy of Sciences,\\
	and Beijing National Laboratory for Condensed Matter Physics, Beijing 100190, China}

\author{Shuo Li}
\affiliation{Institute of Physics, Chinese Academy of Sciences,\\
	and Beijing National Laboratory for Condensed Matter Physics, Beijing 100190, China}

\author{Aifang Fang}
\affiliation{School of Physics and Astronomy, Beijing Normal University, Beijing 100875, China}
\affiliation{Key Laboratory of Multiscale Spin Physics, Ministry of Education, Beijing Normal University, Beijing 100875, China}

\author{Jie Yang}
\affiliation{Institute of Physics, Chinese Academy of Sciences,\\
	and Beijing National Laboratory for Condensed Matter Physics, Beijing 100190, China}

\author{Yanpeng Qi}
\email{qiyp@shanghaitech.edu.cn}
\affiliation{State Key Laboratory of Quantum Functional Materials, School of Physical Science and Technology, ShanghaiTech University, Shanghai 201210, China}
\affiliation{ShanghaiTech Laboratory for Topological Physics, ShanghaiTech University, Shanghai 201210, China}
\affiliation{Shanghai Key Laboratory of High-resolution Electron Microscopy, ShanghaiTech University, Shanghai 201210, China}

\author{Junjie Zhang}
\email{junjie@sdu.edu.cn}
\affiliation{State Key Laboratory of Crystal Materials, Institute of Crystal Materials, Shandong University, Jinan, Shandong 250100, China}

\author{Rui Zhou}
\email{rzhou@iphy.ac.cn}
\affiliation{Institute of Physics, Chinese Academy of Sciences,\\
 and Beijing National Laboratory for Condensed Matter Physics, Beijing 100190, China}
\affiliation{School of Physical Sciences, University of Chinese Academy of Sciences, Beijing 100190, China}

\date{\today}% It is always \today, today,

\begin{abstract}
{The discovery of superconducting transitions in pressurized La$_3$Ni$_2$O$_{7}$ and La$_4$Ni$_3$O$_{10}$ has highlighted the pivotal role of density wave (DW) orders in nickelate superconductors. To gain a comprehensive understanding of the superconducting state, it is essential to elucidate the nature of the DW order. In this study, we utilized $^{139}$La nuclear quadrupole resonance (NQR) to investigate the charge density wave (CDW) and spin density wave (SDW) orders in both single-crystal and polycrystalline La$_4$Ni$_3$O$_{10}$. Near $T_{\rm{DW}} \approx 133$ K, an abrupt change in both the linewidth and frequency of the La(2) site in the single-crystal sample provides compelling evidence for a first-order-like phase transition. The pronounced broadening of the NQR lines indicates the incommensurate nature of the DW order. Furthermore, the spin-lattice relaxation rate divided by temperature 1/$T_1$$T$ exhibits a strong enhancement at $T_{\rm{DW}}$, indicating the strong spin fluctuations above the first-order DW transition. These observations suggest an intricate interplay between incommensurate CDW and SDW orders. Our findings offer critical insights into the microscopic mechanisms of the DW state in La$_4$Ni$_3$O$_{10}$ and establish an essential framework for exploring the interplay between DW and superconducting phases in nickelate superconductors.
}
\end{abstract}


%\keywords{High-temperature superconductors, Nuclear magnetic resonance, pair density wave, Pseudogap}

\maketitle

\section{Introduction}
The recent discovery of superconductivity in nickelates, particularly the breakthrough in pressurized Ruddlesden-Popper (RP) phase La$_3$Ni$_2$O$_7$\cite{Sun2023,Wang2024b,Lifeiyu2025} and La$_4$Ni$_3$O$_{10}$\cite{Zhu2024,Li2024wen,ZhangMX2025,LIFEIYU2025}, has presented a new opportunity for studying the superconducting pairing mechanism of high-$T_{\rm c}$ superconductors. Although RP-phase nickelates are non-superconducting at ambient pressure, magnetic orders were observed in different nickelates\cite{Kobayashi1996,Ram1986,Sreedhar1994,Wu2001,Luo2025,shulei2024,wutao025,fengdonglai} -- a feature shared with both cuprate\cite{cuprateReview} and iron-based superconductors\cite{FeReview}. Therefore, understanding the emergence of magnetic ordered states is of great significance for understanding superconductivity{\cite{HXLi2017,Li2024,Leonov2024,Wang2024,LaBollita2024}}. 
Among various nickelates, La$_4$Ni$_3$O$_{10}$ has attracted significant  attention due to its high superconducting volume fraction\cite{Zhu2024}, well-defined interlayer stacking ordering and intertwined charge- and spin- density wave (DW) orders\cite{Zhang2020}. These characteristics not only make La$_4$Ni$_3$O$_{10}$ an important platform for investigating DW orders but also facilitate the exploration of the microscopic origin of unconventional superconductivity. 

At ambient pressure, La$_4$Ni$_3$O$_{10}$ crystallizes in the monoclinic P$2_1$/a space group (Fig. \ref{fig:crystalstructure}(a))\cite{Zhu2024}, with NiO$_6$ octahedra tilted away from the $c$-axis. Early electrical and magnetic transport studies\cite{Kobayashi1996,Ram1986,Sreedhar1994} have provided evidence for the potential emergence of DW orders in La$_4$Ni$_3$O$_{10}$ near 140 K. Subsequently, through low-temperature X-ray diffraction (XRD) and neutron scattering experiments on single crystals, both charge- and spin-density wave orders have been observed in La$_4$Ni$_3$O$_{10}$\cite{Zhang2020}. Under high pressures, the suppression of the DW order and the emergence of the superconducting phase occur nearly simultaneously in the vicinity of the structural transition\cite{Zhu2024}, suggesting a potential competitive interplay among these electronic orders. Later, various microscopic techniques have been utilized to obtain more microscopic information regarding the DW state. Ultrafast spectroscopy has uncovered a weaker correlation in La$_4$Ni$_3$O$_{10}$ compared to that in the bilayer nickelate La$_3$Ni$_2$O$_{7}$\cite{Li2025b,Xu2025}. Nevertheless, the DW order still strongly competes with superconductivity \cite{Xu2025b}.  Scanning tunneling microscopy (STM) measurements\cite{Li2025} have uncovered an incommensurate unidirectional CDW in La$_4$Ni$_3$O$_{10}$, which is consistent with previous XRD findings\cite{Zhang2020}. Recent muon spin relaxation ($\mu$SR) measurements\cite{Cao2025,Khasanov2025} suggest that the SDW order is also incommensurate, which seems to be consistent with previous neutron scattering measurements\cite{Zhang2020}. Nevertheless, a second SDW order was observed in polycrystalline samples by $\mu$SR measurements\cite{Cao2025,Khasanov2025}, and this has not been observed in any other measurements to date. Simultaneously, the question of whether the DW orders are predominantly governed by charge, spin, or an intertwined order remains unresolved, and their origin and microscopic mechanisms necessitate systematic investigation. Thus, a local probe is of crucial importance for clarifying the microscopic mechanisms underlying DW orders.

%High pressure studies demonstrated that La$_4$Ni$_3$O$_{10}$ undergoes a pressure-induced structural transition to the tetragonal I4/mmm phase\textsuperscript{\cite{SJPD36FAC8D9999EAE57A0BA0EF0E0AC3C15,cpl_41_1_017401}}, accompanied by a progressive increase of Ni–O–Ni bond angles toward 180°. 

%Intriguingly, , direct evidence of their intertwined coexistence remains lacking.

The nuclear quadrupole resonance (NQR) exhibits high sensitivity to local electric field gradients (EFG) and internal magnetic fields associated with CDW and SDW formations. Owing to the use of polycrystalline samples and limitations in sample quality, previous nuclear magnetic resonance (NMR) and NQR studies\cite{Kakoi2024,Fukamachi2001b} on La$_4$Ni$_3$O$_{10}$ have not provided detailed spectroscopic information regarding the DW transition, particularly concerning SDW order and related magnetic fluctuations. In this article, we utilize $^{139}$La(2) NQR to explore DW orders in polycrystalline and single-crystal La$_4$Ni$_3$O$_{10}$ under ambient pressure. Below $T_{\rm{DW}} \approx 133$ K, the sudden broadening of the linewidth and frequency shifts, particularly in the ±5/2 $\leftrightarrow$ ±7/2 transition, reveal a first-order-like DW transition. The significant broadening of the NQR lines suggests the incommensurate nature of the DW order. Moreover, the temperature-dependent spin-lattice relaxation rate 1/$T_{1}$$T$ shows a notable enhancement at $T_{\rm{DW}}$, suggesting the onset of strong spin fluctuations above the first-order DW transition. These findings imply that the first-order transition is likely induced by the CDW, while the spin fluctuations stem from the SDW, indicating a complex interplay between incommensurate CDW and SDW orders. These results not only elucidate the microscopic mechanisms underlying the DW state in La$_4$Ni$_{3}$O$_{10}$ but also establish a crucial framework for investigating the interaction between DW and superconducting phases in nickel-based superconductors.

\section{NQR Methods}

The polycrystalline and single-crystal samples of La$_4$Ni$_3$O$_{10}$ were synthesized via the sol-gel method \cite{ZHANG2024147} and flux growth method \cite{Fyl}, respectively. The $^{139}$La NQR spectra were acquired by sweeping the frequency point by point and integrating the spin-echo intensity in the absence of any applied static magnetic fields. The amounts of the polycrystalline and single-crystal samples employed for NQR measurement were approximately 600 mg and 80 mg, respectively. Given that the principal axis of $^{139}$La is aligned along the $c$-axis, the single-crystal samples were oriented along the $c$-axis, which could ensure that the radio frequency field $H_1$ was within the $ab$ plane. This enabled us to use a much smaller coil for NQR measurements on single crystal samples as compared to polycrystalline samples. 
During NQR experiments, a much larger ringing signal of the polycrystalline sample was observed at low frequencies. As a result, an interval of $\tau$ = 120 $\mu$s between the $\pi/$2 and $\pi$ pulses was used for the polycrystalline samples at low frequencies, while only an interval of 50 $\mu$s was used for single crystal samples. The use of a longer $\tau$ will lead to the loss of the NQR signal at low frequencies for polycrystalline samples. The spin-lattice relaxation time $T_1$ was measured using the saturation-recovery method. 
The spin-lattice relaxation time $T_1$ and the stretched-exponential factor $\beta$ are achieved by fitting the relaxation curves via the stretched exponential formula\cite{Chepin1991}: $(M_0 - M(t))/M_0 = {\frac{3}{22}}{e^{-21(\frac{t}{T_1})^{\beta}}}+{\frac{50}{77}}{e^{-10(\frac{t}{T_1})^{\beta}}}+{\frac{3}{14}}{e^{-3(\frac{t}{T_1})^{\beta}}}$. Because $\eta$ is close to zero, we ignore the influence of $\eta$ on the recovery curve. 

\section{Results and Discussions}

\begin{figure}[htbp]
	\centering
	\includegraphics[width=1.0\linewidth]{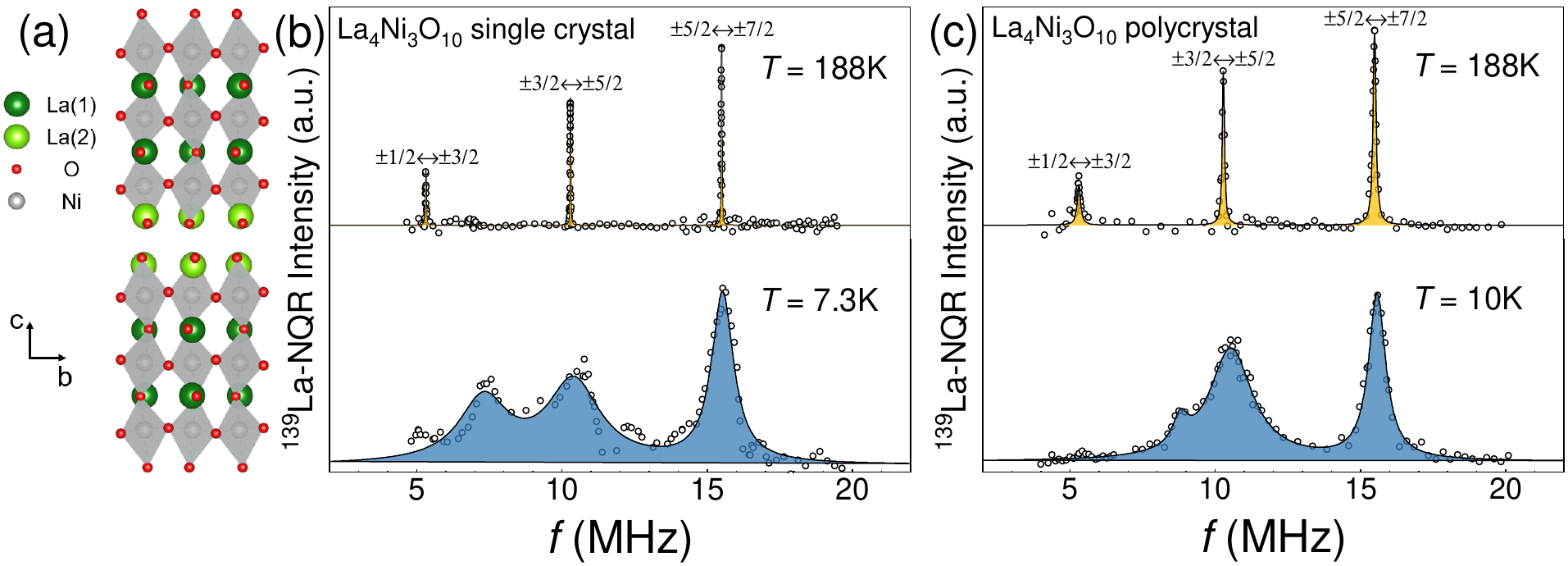}
	\caption{%Crystal structure and $^{139}$La(2) NQR spectra of La$_4$Ni$_3$O$_{10}$ at various temperatures.
(a) Crystal structure of La$_4$Ni$_3$O$_{10}$. Light green atoms denote La(2) atoms, while dark green atoms denote La(1) atoms. 
(b-c) The $^{139}$La NQR spectra of La$_4$Ni$_3$O$_{10}$ single-crystal (b) and polycrystalline (c) samples in the normal state and the DW state, respectively. The yellow and blue peaks correspond to the signals of La$_4$Ni$_3$O$_{10}$ before and after the transition, respectively. All lines are fitted using Lorentz functions, which serve as visual guides.}
	\label{fig:crystalstructure}
\end{figure}

La atoms occupy two distinct crystallographic sites as shown in Fig. \ref{fig:crystalstructure}(a). Specifically, La(1) is positioned within the NiO$_2$ bilayers, while La(2) is located outside the NiO$_2$ trilayers. The quadrupole frequency is defined as $v_q = \frac{3e^{2}qQ}{2I(2I - 1)h}$, where $eq$ represents the EFG, $Q$ is the nuclear quadrupole moment, $I$ denotes the nuclear spin quantum number, and $h$ is the Planck's constant. For the $^{139}$La nucleus with a spin of 7/2, the NQR spectrum is expected to exhibit three resonance lines corresponding to the ±1/2 $\leftrightarrow$ ±3/2, ±3/2 $\leftrightarrow$ ±5/2, and ±5/2 $\leftrightarrow$ ±7/2 transitions. Thus, a total of six lines are anticipated to be observable in the $^{139}$La NQR spectrum of La$_4$Ni$_3$O$_{10}$. However, the La(1) site exhibits a significantly lower $v_q$ (2.2 MHz) and a longer $T_1$ \cite{Kakoi2024,Fukamachi2001b} compared to that of the La(2) site. As a result, at 188 K, only three characteristic NQR lines from the La(2) sites are observed in single-crystal and polycrystalline samples, respectively (see Fig. \ref{fig:crystalstructure}(b) and \ref{fig:crystalstructure}(c), top panels). Compared with the La$_4$Ni$_3$O$_{10}$ polycrystalline samples, the NQR linewidth of the single crystal samples is much smaller at 188 K, indicating that the single-crystal samples have better crystal quality. In other words, the single-crystal samples contain fewer disorders or oxygen vacancies. 
 
\begin{figure}[htbp]
	\centering
	\includegraphics[width=1.0\linewidth]{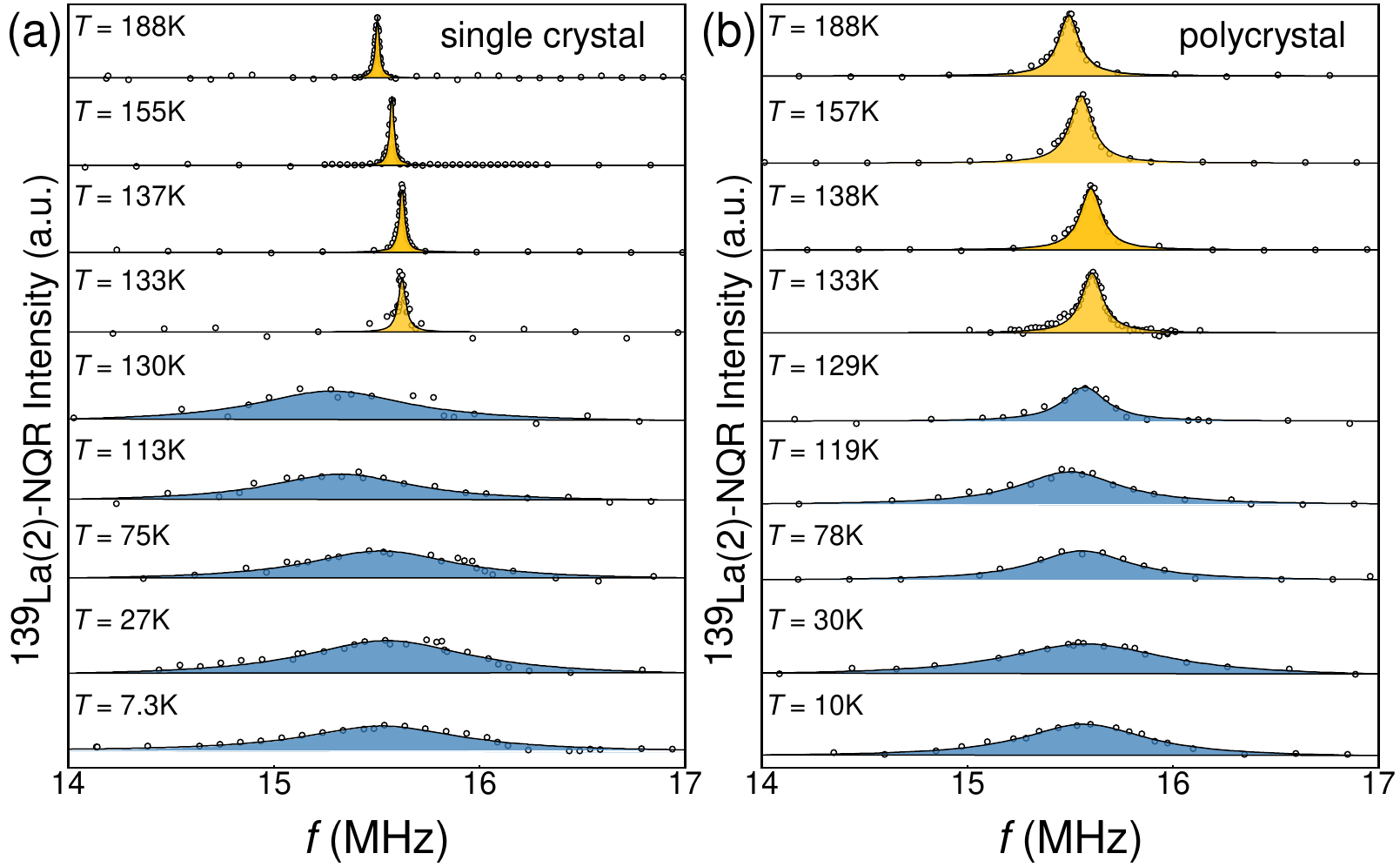}
	\caption{The temperature-dependent $^{139}$La(2) NQR spectra of La$_4$Ni$_3$O$_{10}$ single-crystal (a) and polycrystal (b) samples. The yellow and blue peaks represent the ±5/2 $\leftrightarrow$ ±7/2 transition lines of La$_4$Ni$_3$O$_{10}$ before and after the DW transition, respectively. The solid lines represent the Lorentz fit.}
	\label{fig:spectrum}
\end{figure}
 
As the temperature decreases far below the DW transition temperature $T_{\rm DW}$, a notable line broadening is observed in both single-crystal and polycrystal samples (see Fig. \ref{fig:crystalstructure}(b) and \ref{fig:crystalstructure}(c), bottom panels). For both types of samples, only one broad line is observable around the two NQR lines corresponding to the ±5/2 $\leftrightarrow$ ±7/2 and ±3/2 $\leftrightarrow$ ±5/2 transitions. Nevertheless, in the case of single-crystal samples, an additional broad line is seen at lower frequencies. This could be attributed to the fact that a much smaller interval $\tau$ between $\pi$/2 and $\pi$ pulses can be employed for NQR measurements of single-crystal samples. The use of a much longer $\tau$, resulting from the emergence of a larger ringing signal, can cause the loss of the NQR signal at low frequencies in polycrystalline samples. 

\begin{figure}[htbp]
	\centering
	\includegraphics[width=1.0\linewidth]{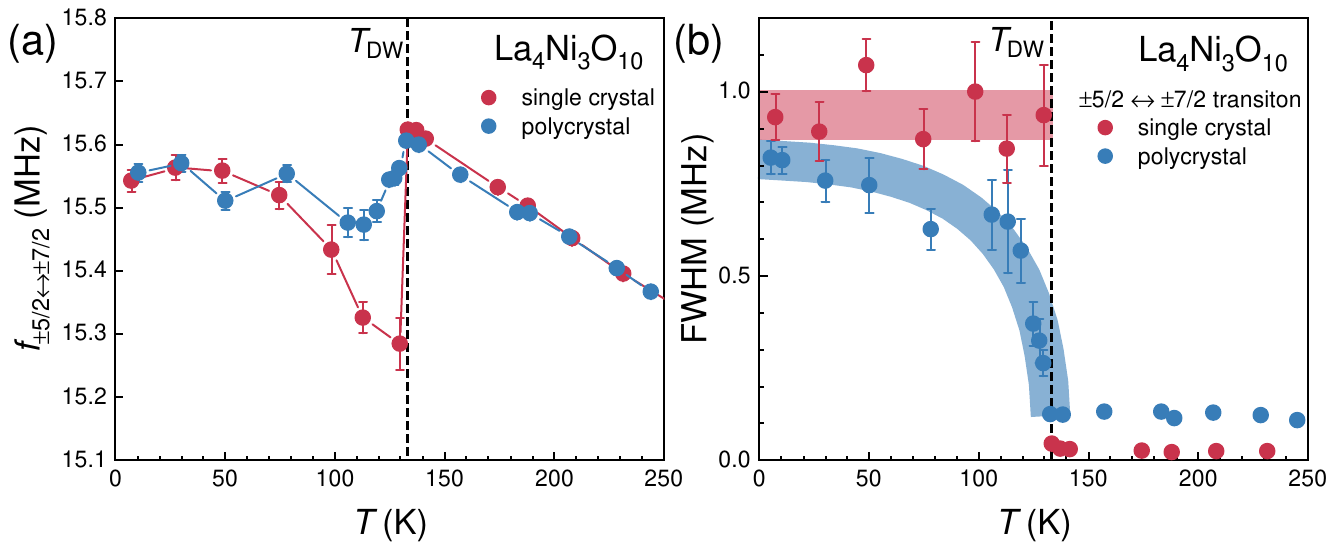}
	\caption{The temperature-dependent NQR frequency (a) and full width at half maximum(FWHM) (b) of the line corresponding to the ±5/2 $\leftrightarrow$ ±7/2 transition in single-crystal(red dots) and polycrystal(blue dots) La$_4$Ni$_3$O$_{10}$. Both of them start to increase below $T_\textrm{DW} \approx$ 133 K marked by the dashed line. The error bar is the s.d. in fitting spectra by the Lorentz function.}
	\label{fig:FWHMandf20250523}
\end{figure}

Figure \ref{fig:spectrum} further shows the temperature dependence of the NQR lines corresponding to the ±5/2 $\leftrightarrow$ ±7/2  transition for both single crystal and polycrystalline samples, respectively. Although a clear broadening appears below $T_{\rm DW}$ for both types of samples, the temperature-dependent behaviors just below $T_{\rm DW}$ are not identical. 
In the case of single-crystal samples, the line broadening and shift at the ±5/2 $\leftrightarrow$ ±7/2 transition occur abruptly below the DW transition temperature $T_{\rm{DW}} \approx$ 133 K (as shown in Fig. \ref{fig:spectrum}(a)). However, for polycrystalline samples below $T_{\rm{DW}} \approx$ 133 K, a gradual change in linewidth and frequency is observed. 
Figure \ref{fig:FWHMandf20250523} summarizes the temperature dependence of the frequency and full width at half maximum (FWHM) of the NQR line corresponding to the ±5/2 $\leftrightarrow$ ±7/2 transition. It is readily apparent that the DW transition in the single-crystal sample exhibits characteristics reminiscent of a first-order transition. 
%Both the frequency and FWHM of the NQR line experience abrupt changes below $T_{\rm{DW}} \approx$ 133 K, which is in contrast to the polycrystalline sample. 
Considering the better quality of single crystal samples, the more abrupt change below $T_{\rm{DW}}$ is more likely associated with the intrinsic characteristics of La$_4$Ni$_3$O$_{10}$. A greater degree of disorder or a larger distribution of oxygen vacancies in the sample would render the transformation more gradual. Similar behavior was observed through specific heat measurements, where a smaller jump of the specific heat at the transition was seen in polycrystalline samples\cite{Khasanov2025}. 
Our observation in La$_4$Ni$_3$O$_{10}$ differs from previous NQR results in La$_3$Ni$_2$O$_7$, where the DW phase transition was identified as second-order\cite{Luo2025}. Especially, we note that previous $\mu$SR studies on single-crystal samples indicated a second-order phase transition\cite{shulei2024}, suggesting that La$_3$Ni$_2$O$_7$ and La$_4$Ni$_3$O$_{10}$ likely display distinct behaviors. However, earlier NQR studies used polycrystalline samples\cite{Kakoi2024,Fukamachi2001b}, and the spectra near the transition may resemble those of polycrystalline La$_4$Ni$_3$O$_{10}$, which are influenced by sample quality. To conclusively determine the nature of the transition, it would be essential to conduct similar NQR measurements on high-quality single-crystal samples in the future. 
Additionally, for both single crystal and polycrystal samples, no abrupt increases in FWHM was observed below $T \approx$ 90 K, at which a second SDW transition was observed by recent $\mu$SR measurements\cite{Cao2025,Khasanov2025}. Given that the La(2) site is only sensitive to the magnetic moments of the outer NiO$_2$ planes, it is plausible that the change observed by $\mu$SR measurements is associated with the inner NiO$_2$ planes. Further NMR studies on the La(1) site are necessary to clarify this issue. 

Next we discuss the origin of the changes in the NQR spectra in the DW state. Firstly, unlike La$_3$Ni$_2$O$_{7}$, we did not observe splitting on each NQR line, especially the NQR line corresponding to the ±5/2 $\leftrightarrow$ ±7/2 transition\cite{Luo2025}. This indicates that the DW order in this sample is incommensurate and different from that of La$_3$Ni$_2$O$_{7}$. In this case, it would only cause broadening of the NQR spectrum. Meanwhile, we did not observe the double-horn structure of one-dimensional incommensurate peaks\cite{Blinc2002}, indicating that the incommensurate modulation is most likely to be two-dimensional. In principle, the broadening of the NQR spectra arises from two distinct contributions: one is due to changes in the local EFG caused by charge modulation, and the other originates from the formation of an internal magnetic field $B_{\rm int}$. 
If the internal magnetic field at the La(2) site aligns along the $c$-axis, both the ±3/2 $\leftrightarrow$ ±5/2 and ±5/2 $\leftrightarrow$ ±7/2 transitions would shift and broaden identically\cite{Luo2025}.  However, the broadening of the NQR lines corresponding to the ±1/2 $\leftrightarrow$ ±3/2 and ±3/2 $\leftrightarrow$ ±5/2 transitions are significantly larger than that of the ±5/2 $\leftrightarrow$ ±7/2 transition in both single-crystal and polycrystalline samples at low temperatures (see Fig. \ref{fig:crystalstructure}(b) and \ref{fig:crystalstructure}(c), bottom panels). Meanwhile the two low frequency lines also shift to higher frequencies. 
If the internal magnetic field at the La(2) site aligns perpendicular to the $c$-axis, the NQR lines corresponding to the ±1/2 $\leftrightarrow$ ±3/2 and ±3/2 $\leftrightarrow$ ±5/2 transitions should shift to higher frequencies and broaden a lot as indicated by previous NQR calculations\cite{Luo2025}. This seems to in agreement with our experimental results. However, in this case, the internal magnetic field would barely affect the NQR line corresponding to the ±5/2 $\leftrightarrow$ ±7/2 transition, which can not explain the large increase of linewidth of this NQR line. Besides this, we also note that the change of NQR lines cannot be solely attributed to quadrupole broadening caused by charge modulation too. In the case of quadrupole broadening, one would expect the broadening of each NQR lines is proportional to the NQR frequencies\cite{Luo2025}, which is also inconsistent with our observation. Taken all together, our results implies that the change of the NQR spectra is not related to only one kind of DW. Instead, both charge and spin density waves develop below $T_{\rm{DW}}$, which is consistent with previous neutron scattering studies\cite{Zhang2020}.

%the ±3/2 $\leftrightarrow$ ±5/2 transition to be much smaller than that of the ±5/2 $\leftrightarrow$ ±7/2 transition, as the broadening is proportional to the NQR frequency.

%both the ±3/2 $\leftrightarrow$ ±5/2 and ±5/2 $\leftrightarrow$ ±7/2 transitions would broaden identically..

%Therefore, the substantial broadening of the ±3/2 $\leftrightarrow$ ±5/2 transition is most likely caused by an internal magnetic field that is oriented perpendicular to the $c$-axis. 

%Additionally, the abrupt change in the NQR frequency of the ±5/2 $\leftrightarrow$ ±7/2 transition just below $T_{\rm{DW}}$ suggests a significant modification of the EFG. This observation implies that the broadening of the ±5/2 $\leftrightarrow$ ±7/2 transition is primarily due to the formation of a CDW. 

\begin{figure}[htbp]
	\centering
	\includegraphics[width=1.0\linewidth]{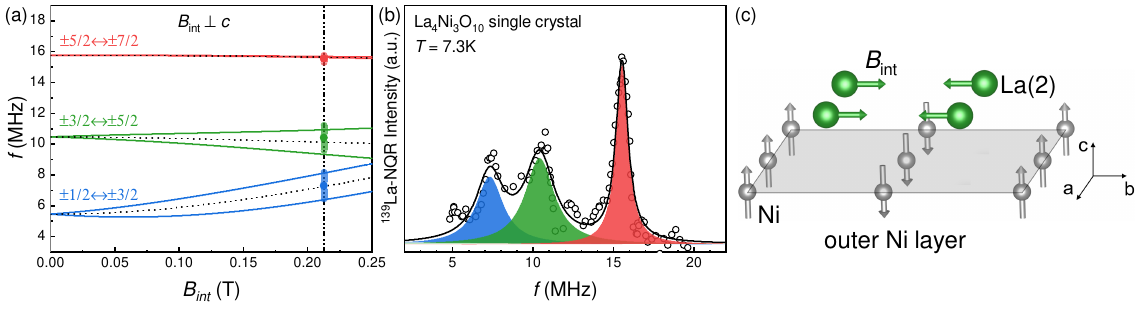}
	\caption{(a) Theoretical simulation of the NQR frequencies at the La(2) site with an internal magnetic field perpendicular to the $c$-axis. The red, green, and blue curves represent the field-dependent shifts of the ±5/2 $\leftrightarrow$ ±7/2, ±3/2 $\leftrightarrow$ ±5/2 and ±1/2 $\leftrightarrow$ ±3/2 transitions, respectively. If the SDW order is commensurate, distinct line splitting should be discerned for NQR lines corresponding to the ±5/2 $\leftrightarrow$ ±7/2 and ±3/2 $\leftrightarrow$ ±5/2 transitions. Nevertheless, in the DW state, only broad lines are observed owing to the incommensurate spin and charge modulation. The black dashed line depicts their average frequencies in the DW state. Solid circles represent the frequencies of three NQR lines derived from the experimental results presented in panel (b). The color bars represent the FWHM contributed by the internal magnetic field, which is proportional to the splitting of NQR lines due to SDW.  (b) Spectrum of La$_4$Ni$_3$O$_{10}$ single crystal at 7.3 K. The solid lines denote Lorentz fits. The resolved three peaks correspond to the ±5/2 $\leftrightarrow$ ±7/2(red), ±3/2 $\leftrightarrow$ ±5/2(green) and ±1/2 $\leftrightarrow$ ±3/2(blue) transitions. The $v_q$ and $\eta$ are taken as 5.27 MHz and 0.1 for the calculation, respectively. (c) Schematic diagram of the internal magnetic field at the La(2) site within La$_4$Ni$_3$O$_{10}$. The grey and green arrows represent the Ni magnetic moments and the internal magnetic field $B_{int}$, respectively.}
	\label{fig:simulation}
\end{figure}

%\begin{table}[h!]
%	\begin{center}
%		\caption{Contributions of CDW and SDW orders on NQR lines spreading.}
%		\label{tab:cdw_sdw_contributions}
%		\begin{tabular}{|c|c|c|c|}  % 四列对齐：左|中|中|右
%			\hline % 顶部边框
%			\textbf{Transition} & \textbf{$W_{\text{total}}$ (MHz)} & \textbf{$W_{\text{CDW}}$ (MHz)} & \textbf{$W_{\text{SDW}}$ (MHz)} \\
%			\hline % 表头下方边框
%			±5/2 $\leftrightarrow$ ±7/2 & 0.90 & 0.90 & 0.00 \\ 
%			\hline % 行间分隔线
%			±3/2 $\leftrightarrow$ ±5/2 & 2.20 & 0.60 & 1.60 \\   
%			\hline % 行间分隔线
%			±1/2 $\leftrightarrow$ ±3/2 & 2.04 & 0.30 & 1.74 \\  
%			\hline % 底部边框
%		\end{tabular}
%	    \begin{minipage}{\textwidth}
%	    	\vspace{4mm} % 添加一点垂直间距
%	    	\footnotesize % 使用小号字体
%	    	\textit{Note:} $W_{\text{total}}$ represents the total linewidth broadening, $W_{\text{CDW}}$ denotes the contribution from the CDW order and $W_{\text{SDW}}$ indicates the contribution from the SDW order. Measurements were performed at 7.3 K for La$_4$Ni$_3$O$_{10}$ single crystal.
%	    \end{minipage}
%	\end{center}
%\end{table}

As we noted that the two NQR lines corresponding to ±1/2 $\leftrightarrow$ ±3/2 and ±3/2 $\leftrightarrow$ ±5/2 exhibit a clear shift towards higher frequencies in Fig. \ref{fig:simulation}(b), we assume that an internal magnetic field $B_{\rm int}$ at the La(2) site is perpendicular to the $c$-axis. To simulate the experimentally obtained NQR spectrum, as shown in Fig. \ref{fig:simulation}(a), we assume that the broadening caused by charge modulations is $W_{\rm CDW}$. In the DW state, the spectrum is broadened nearly symmetrically and follows a Lorentzian distribution. Since the FWHM of the convolution of two Lorentzian functions is the sum of their individual FWHM, the total broadening of each NQR spectrum can be considered to consist of two components: $W^i$ = $W^{±i \leftrightarrow ±(i+1)}_{\rm CDW}$ + $W^{±i \leftrightarrow ±(i+1)}_{\rm SDW}$ ($i$ = 1/2, 3/2, 5/2). Among these, $W^{±i \leftrightarrow ±(i+1)}_{\rm CDW}$ is proportional to the frequency of the NQR line, such that $W^{±1/2 \leftrightarrow ±3/2}_{\rm CDW}$:$W^{±3/2 \leftrightarrow ±5/2}_{\rm CDW}$:$W^{±5/2 \leftrightarrow ±7/2}_{\rm CDW}$ = 1:2:3.  $W^{±i \leftrightarrow ±(i+1)}_{\rm SDW}$ is proportional to the splitting of the NQR spectrum induced by the internal magnetic field $B_{int}$. The shift of the NQR spectrum can also be attributed to $B_{\rm int}$. As shown in Fig. \ref{fig:simulation}(a), we found that when $B_{\rm int}$ $\sim$ 210 mT and $W^{1/2}_{\rm CDW}$ = 0.3 MHz, both the shift and broadening of the NQR spectrum are consistent with the experimental results. Our simulation results also suggest, the pronounced NQR line shift and broadening in La$_4$Ni$_3$O$_{10}$ in the DW state originate from collective modulation by incommensurate CDW and SDW orders. We further note that the internal magnetic field arising from the magnetic moment of Ni is significantly larger than that in La$_3$Ni$_2$O$_{7}$ ($B_{\rm int}$ $\sim$ 5mT)\cite{Luo2025}. Our results suggest that the magnetic moment of Ni on the outer planes is much larger in La$_4$Ni$_3$O$_{10}$ than that in La$_3$Ni$_2$O$_{7}$\cite{Luo2025}. Moreover, we note that previous neutron scattering experiments have indicated a stripe-type antiferromagnetic order\cite{Zhang2020}. Regarding the orientation of the Ni magnetic moments, one possibility is that they lie within the $ab$-plane, as observed in Fe-based superconductors\cite{FeReview}. However, due to coupling with the four neighboring Ni moments, this would result in an internal magnetic field at the La(2) site aligned along the $c$-axis\cite{Lizheng2012}, which is inconsistent with our experimental observations. Alternatively, as proposed by a prior theoretical study\cite{LaBollita2024} (see Fig. \ref{fig:simulation}(c)), the Ni magnetic moments on the outer Ni layer may be oriented along the $c$-axis. In this case, the internal magnetic field at the La(2) site in La$_4$Ni$_3$O$_{10}$ would be perpendicular to the $c$-axis, in agreement with our simulation results (Fig. \ref{fig:simulation}(c)). Thus, our findings support the interpretation that the Ni magnetic moments on the outer layer are aligned along the $c$-axis in La$_4$Ni$_3$O$_{10}$, which is similar to the observation in La$_3$Ni$_2$O$_{7}$\cite{wutao025}.

\begin{figure}[htbp]
	\centering
	\includegraphics[width=0.5\linewidth]{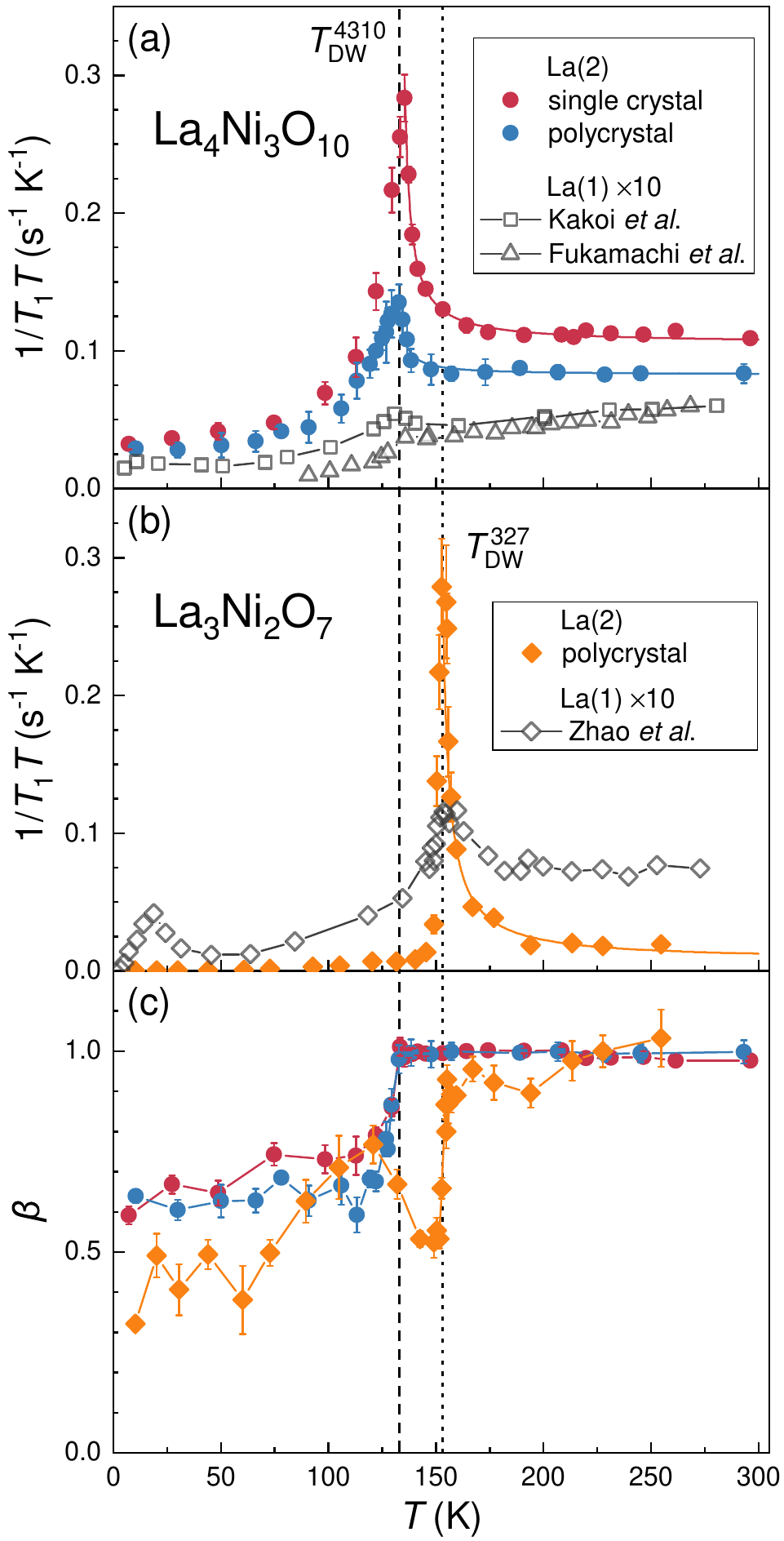}
	\caption{(a) Red dots and blue symbols represent the 1/$T_1T$ measured at the La(2) site in La$_4$Ni$_3$O$_{10}$ single crystal and polycrystal, respectively. Grey squares and triangles represent the 1/$T_1$$T$ measured at the La(1) site reported  by  T. Fukamachi et al.\cite{Fukamachi2001b} and M. Kakoi et al.\cite{Kakoi2024}, respectively. (b) Orange and grey diamonds represent the 1/$T_1T$ measured at the La(2) and La(1) sites by J. Luo et al.\cite{Luo2025} and D. Zhao et al.\cite{wutao025}, respectively. (c) Temperature dependence of the $\beta$ of La$_3$Ni$_2$O$_7$(orange diamonds)\cite{Luo2025}, La$_4$Ni$_3$O$_{10}$ single crystal(red circles) and polycrystal(blue circles).
 The error bars of 1/$T_1T$ and $\beta$ are the s.d. in fitting the recovery curve. The dashed and dotted lines represent $T_\textrm{DW}$ in La$_4$Ni$_3$O$_{10}$ and La$_3$Ni$_2$O$_7$, respectively. Note that all the 1/$T_1T$ values for the La(1) site have been scaled by a factor of 10.}
	\label{fig:T1_and_beta}
\end{figure}

We further investigated the DW transition through spin-lattice relaxation rate measurements. Similar to the case of La$_3$Ni$_2$O$_7$\cite{Luo2025}, a pronounced peak is observed in the temperature-dependent 1/$T_1$$T$  at the transition temperature $T_{\rm{DW}}$ for the single-crystal sample as shown in Fig. \ref{fig:T1_and_beta}. For both La$_3$Ni$_2$O$_7$ and La$_4$Ni$_3$O$_{10}$, the stretched-exponential factor $\beta$ decreases rapidly below $T_{\rm{DW}}$ shown in Fig. \ref{fig:T1_and_beta}(c), resulting from the spin and charge modulation in the DW order. In comparison with the single-crystal samples, the polycrystal samples exhibit a much weaker enhancement, further highlighting the superior quality of the single-crystal samples.  Similar to La$_3$Ni$_2$O$_7$, the strong enhancement of 1/$T_1$$T$ is most likely attributable to spin fluctuations. However, considering that the DW transition in La$_4$Ni$_3$O$_{10}$ is of the first order, spin fluctuations associated with the DW order are expected to be negligible above $T_{\rm{DW}}$. In La$_3$Ni$_2$O$_7$, where the transition is second order, such fluctuations are naturally more prominent. The nearly identical enhancement of 1/$T_1$$T$  in La$_4$Ni$_3$O$_{10}$, despite its first-order transition nature, appears contradictory. A plausible explanation is that the first-order transition is driven by the charge density wave\cite{Norman2025}, while the strong enhancement of 1/$T_1$$T$ originates from spin fluctuations, akin to what has been observed in the kagome metal CsV$_3$Sb$_5$\cite{Luo2022,Feng2023}. Additionally, we note that the temperature-dependent 1/$T_1$$T$ results reported by M. Kakoi  et al. (Fig. \ref{fig:T1_and_beta}(a))\cite{Fukamachi2001b} exhibit only a minor peak. This relatively small enhancement of 1/$T_1$$T$ at $T_{\rm{DW}}$ may be attributed to sample quality. In particular, T. Fukamachi et al. observed no peak in the temperature dependence of 1/$T_1$$T$ at $T_{\rm{DW}}$\cite{Kakoi2024}. However, it is also noteworthy that in La$_3$Ni$_2$O$_7$, the enhancement of 1/$T_1$$T$ at the La(1) site is indeed much smaller than that at the La(2) site (Fig. \ref{fig:T1_and_beta}(b)). Moreover, in both La$_3$Ni$_2$O$_7$ and La$_4$Ni$_3$O$_{10}$, the value of 1/$T_1$$T$ at the La(1) site is much smaller than that at the La(2) site even at high temperatures, suggesting weaker hyperfine coupling with the Ni magnetic moments at La(1) site. Alternatively, in La$_4$Ni$_3$O$_{10}$, the magnetic moments of Ni in the middle layer may be oriented opposite to those in the outer layers\cite{LaBollita2024}, resulting in canceling contributions from the two Ni layers at the La(1) site and thus weaker magnetic fluctuations there.
 %In the end, we also note that a Landau theory approach treats the SDW and CDW as coupled order parameters within a universal phase diagram\cite{Norman2025}. The free energy includes a linear-quadratic coupling term, in which the SDW drives this transition. Nesting at the SDW wave vector supports this scenario, while the CDW appears as the primary order parameter due to strong coupling near the first-order transition region.
%These observations suggest that CDW and SDW orders are intertwined. 

In comparison with La$_3$Ni$_2$O$_7$\cite{Luo2025}, we observe that 1/$T_1$$T$ in La$_4$Ni$_3$O$_{10}$ retains a notable residual value, approximately 20\% of its high-temperature value, even at the lowest temperature. In contrast, 1/$T_1$$T$ in La$_3$Ni$_2$O$_7$ becomes negligibly small at the lowest temperature. These findings indicate that a substantial fraction of the Fermi surface remains ungapped in La$_4$Ni$_3$O$_{10}$, whereas it is nearly fully gapped in La$_3$Ni$_2$O$_7$. Furthermore, the DW order in La$_4$Ni$_3$O$_{10}$ is incommensurate, while it is commensurate in La$_3$Ni$_2$O$_7$. These observations suggest that the emergence of the DW in La$_4$Ni$_3$O$_{10}$ is closely related to Fermi surface nesting\cite{XDu2024}. This nesting not only prevents the complete gapping of the Fermi surface but also gives rise to the incommensurate modulation of the DW orders. Additionally, previous theoretical studies have demonstrated that interlayer coupling plays a pivotal role in the emergence of antiferromagnetism and superconductivity in La$_4$Ni$_3$O$_{10}$\cite{Tian2024,Yang2024,Huo2024}. Specifically, this could explain why the Ni atoms on the outer planes exhibit a much larger magnetic moment in La$_4$Ni$_3$O$_{10}$, yet $T_{\rm SDW}$ is lower compared to that of La$_3$Ni$_2$O$_7$.

\section{Conclusion}

In summary, our $^{139}$La NQR measurements have revealed key microscopic insights into the intertwined CDW and SDW orders in La$_4$Ni$_3$O$_{10}$ at ambient pressure. Below $T_{\rm DW} \approx 133$ K, the abrupt broadening and frequency shift of the NQR lines provide compelling evidence for a first-order-like DW transition. The pronounced broadening of the NQR spectra indicates the incommensurate nature of the DW order. Furthermore, the 1/$T_1$$T$ exhibits a strong enhancement at $T_{\rm DW}$, signifying the strong spin fluctuations originate from the SDW. While the sudden change of the spectra suggests that the first-order transition might be primarily driven by the CDW, highlighting a complex interplay between the CDW and SDW orders. Our findings not only elucidate the microscopic mechanisms underlying the DW state in La$_4$Ni$_3$O$_{10}$ but also provide a crucial framework for understanding the interplay between DW and superconducting phases in nickelate superconductors. 

\begin{acknowledgments}
This work was supported by the National Key Research and Development Projects of China (Grants No. 2023YFA1406103, No. 2024YFA1611302, No. 2024YFA1409200 and No. 2022YFA1403402), the National Natural Science Foundation of China (Grants No. 12374142, No. 12304170 and No. U23A6003), Beijing National Laboratory for Condensed Matter Physics (Grant No. 2024BNLCMPKF005), the CAS Superconducting Research Project under Grant No. [SCZX-01011] and the Science and Technology Commission of Shanghai Municipality (No. 25DZ3008200). J.Z. was supported by the National Natural Science Foundation of China (Grants Nos. 12074219 and 12374457). This work was supported by the Synergetic Extreme Condition User Facility (SECUF, https://cstr.cn/31123.02.SECUF).
\end{acknowledgments}

\section*{DATA AVAILABILITY}
The data that support the finding of this article are not publicly available. The data are available from the authors upon reasonable request.

%\end{thebibliography}

%\vspace{0.5cm}
%\textbf{Author contributions}
%None

%\begin{references}
%\begin{thebibliography}{1}
%\bibliographystyle{unsrt}
%\bibliography{reference}

%\end{thebibliography}
%\end{references}

\end{document}